\documentclass[journal]{IEEEtran}

\ifCLASSINFOpdf
 \usepackage[pdftex]{graphicx}
\usepackage{amsmath}

\ifCLASSOPTIONcompsoc
 \usepackage[caption=false,font=normalsize,labelfont=sf,textfont=sf]{subfig}
\else
 \usepackage[caption=false,font=footnotesize]{subfig}
\fi

\newcommand\tab[1][0.4cm]{\hspace*{#1}}
\usepackage{flushend}
\usepackage{tabularx,booktabs,textcomp}
\newcolumntype{C}{>{\centering\arraybackslash}X} 
\newcolumntype{L}{>{\raggedright\arraybackslash}X} 

\usepackage[noadjust]{cite}

\begin{document}
\title{Optimization of pin GaAs/AlGaAs Heterojunction Nanocone Array Solar Cell based on its Photovoltaic Properties}


\author{Sambuddha~Majumder,
       Sooraj~Ravindran}
\maketitle


\begin{abstract}

In this paper, we have designed and investigated the performance of radial GaAs/AlGaAs pin junction nanocone array solar cells by performing coupled optoelectronic simulations to obtain the most optimal design configuration based on its photovoltaic properties. Each model has been compared with its GaAs shell counterparts for different levels of surface passivations. It has been observed that the nanocones with the AlGaAs shell has a much better performance compared to those having GaAs shell. AlGaAs shell acts as a strong barrier restricting most of the photogeneration to the inner GaAs regions and it also acts as a strong passivation layer, reducing the recombination losses due to surface effects. Further, it is observed that the nanocones achieve their highest photoconversion efficiency when they are sparsely packed, with a constant i-shell thickness of 7-9 nm and have an angle of tilt of 5$^{\circ}$. This enhanced performance is attributed to a more effective and extended photogeneration throughout the nanowire length, a strong overlapping built-in electric field, and lower recombination losses.

\end{abstract}






\section{Introduction}
\IEEEPARstart 
{S}{emiconductor} nanowires (NWs) have become a vast frontier of extensive research in the nanowire industry with their use in LED \cite{guo2010catalyst, qian2005core}, Lasers \cite{duan2003single, saxena2013optically, wilhelm2017broadband}, Photodetectors \cite{dai2014GaAs, wang2013high}, Solar cells \cite{garnett2011nanowire, kayes2008radial, hochbaum2010semiconductor, sandhu2014detailed, sun2011compound, lapierre2013iii} etc. This is attributed to their small structure and excellent light confining abilities. Particularly, nanowire solar cells have been a key area of research in the field of photovoltaics over the past few years. A large number of nanowires have been investigated using materials like Si, GaAs, GaN, AlGaAs, etc., for both axial and radial junction nanowires. Radial junction nanowires, having a junction formed in the radial direction, produce very high-efficiency solar cells with very low material consumption. In contrast to its axial and planar counterparts, radial pin junction serves as an efficient mechanism to decouple light absorption from carrier extraction in the radial direction, leading to lower recombination losses \cite{kupec2009dispersion, kupec2010light, kayes2008radial, yoshimura2013indium}. Further, NWSC can give their optimal performance with very small diameters \cite{wen2011theoretical} and thus can be grown on a large range of lattice-mismatched materials, allowing them to attain efficient multijunction configurations. Extensive work is still going on every day to further optimize each of these designs. Many different optimization strategies have been studied, such as nanowires' structural properties, different surface passivations, nanowire height and diameter, diameter to period ratio, and so on \cite{li2015influence, wu2018optimization, huang2012broadband, wen2011theoretical}.  \\
\tab Recently, a new design methodology has taken over the nanowire solar cell concept, and that is the nanocone solar cells \cite{joyce2013electronic, bai2014one, zhang2018photovoltaic, wang2013high}. It is seen that despite the various methods to optimize the light absorption in the GaAs nanowires, most of the photoabsorption is confined to the top of the nanowires \cite{wu2018optimization, li2015influence}. In contrast, absorption in the depletion region, which spreads throughout the nanowire height, is not sufficient, which leads to only minor improvements to the conversion efficiency. It is observed that providing a taper to the nanowires can increase the absorption properties dramatically. The light absorption in NWs is dominated by resonant modes, which are very closely related to the NW diameter \cite{zhang2018photovoltaic, anttu2010coupling, zhan2014enhanced}. In nanocones, the NW diameter continually changes across the nanowire height, with the top diameter being very smaller than the bottom diameter. Due to this unique geometry, it can support only a few long-wavelength modes in the top, and absorption (particular for long wavelengths) shifts towards the thicker middle regions of the structure \cite{zhang2018photovoltaic}. This even distribution of light throughout the structure increases the effective absorption length, improves the photogeneration, and reduces recombination losses. Also due to the presence of this taper, the NCs acts as a gradual refractive index profile that improves anti-reflection. Therefore, the nanocones must be studied in detail and optimized to improve their performance further. To date, different types of nanocones have been designed and fabricated to investigate the optical properties of the nanocones\cite{bai2014one, joyce2013electronic, soci2008systematic, chuang2011GaAs, wang2013high}. However, very little focus has been given to the photovoltaic properties of these nanocones in solar cells.
Previous work by Zhang $et$ $al.$ \cite{zhang2018photovoltaic} investigates the properties of pin GaAs nanocones; however, GaAs are prone to large surface recombination losses. Previous studies have highlighted the importance of proper radial geometry and good surface passivations for NWSC \cite{li2015influence, lapierre2011numerical}. For GaAs NWs, epitaxial passivation by AlGaAs is an effective method to achieve that \cite{jiang2012long, wu2018optimization}. AlGaAs have lower absorption allowing the photogeneration to be confined to the inner GaAs region and acts as a barrier preventing the carriers generated inside from reaching the surface, thus reducing recombination losses \cite{aaberg2015GaAs, demichel2010impact, chang2012electrical, wu2018optimization}.

\tab In this dissertation, coupled three-dimensional (3D) optical and electrical simulations is done to observe the effect of different nanocone angles on the optical and photovoltaic properties of radial pin junction GaAs and GaAs/AlGaAs nanocones. Four design configurations are chosen for this study, emphasizing the thickness and position of the i-GaAs shell; the best design is then selected and optimized. Finally, a period study is done to observe the effect of dense and sparse packing of the nanocones on their performance. In Section~\ref{sec:mm}, the various electronic and optical properties of the materials and the different design models and methods are discussed. Section~\ref{sec:nas} examines the effect of the nanocone angle on different nanocone properties. Section~\ref{sec:nds} compares the different nanocone design models, and the best model is then decided and optimized. Finally, in Section~\ref{sec:ps} a study of the nanocone period is done.


\section{Modeling and Methods}
\label{sec:mm}
\begin{figure}[t]
        \centering
        \subfloat[]{\label{fig:ua}\includegraphics[width=3.45in]{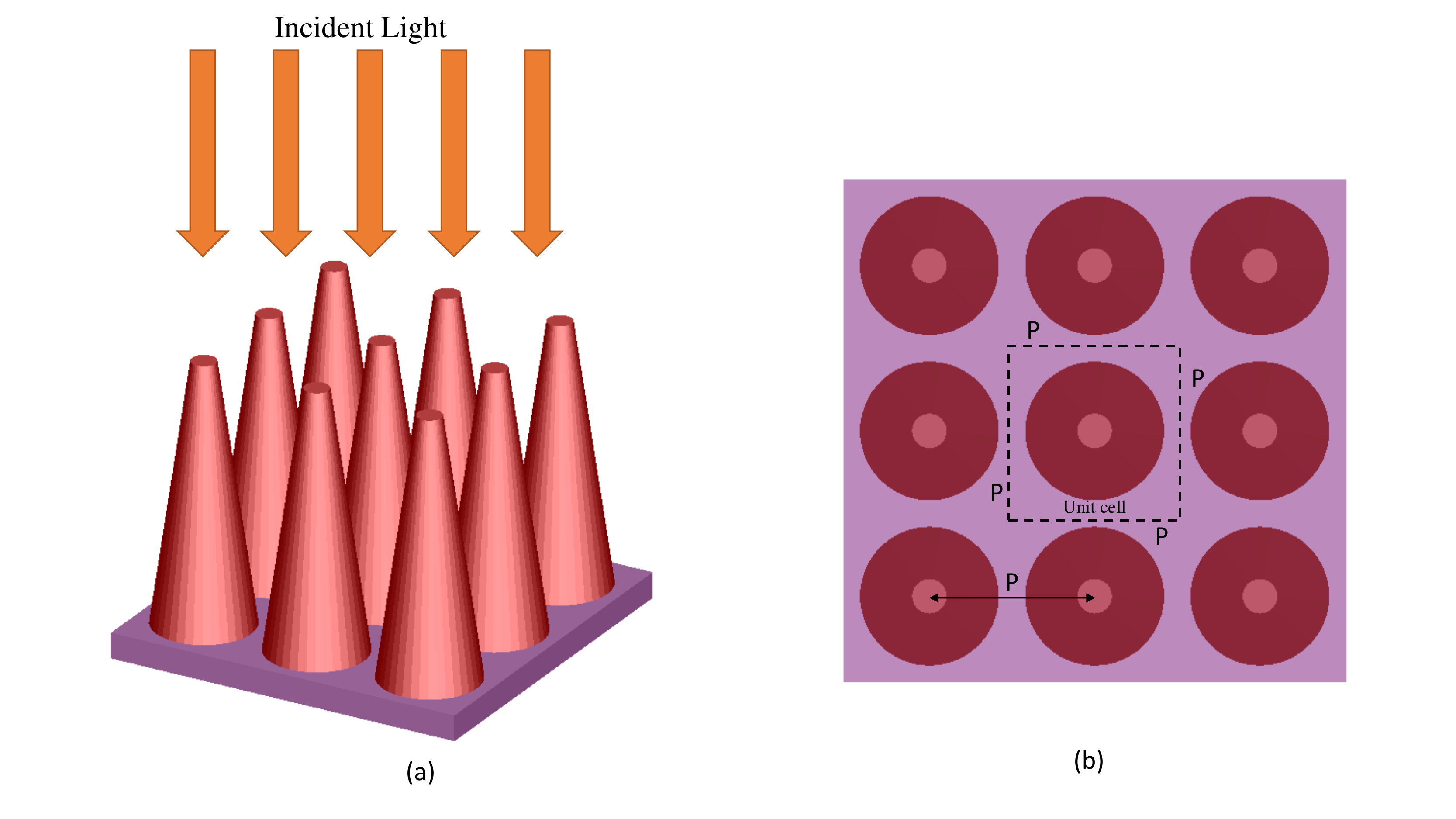}}
        \qquad
        \subfloat[]{\label{fig:ub}\includegraphics[width=3.35in]{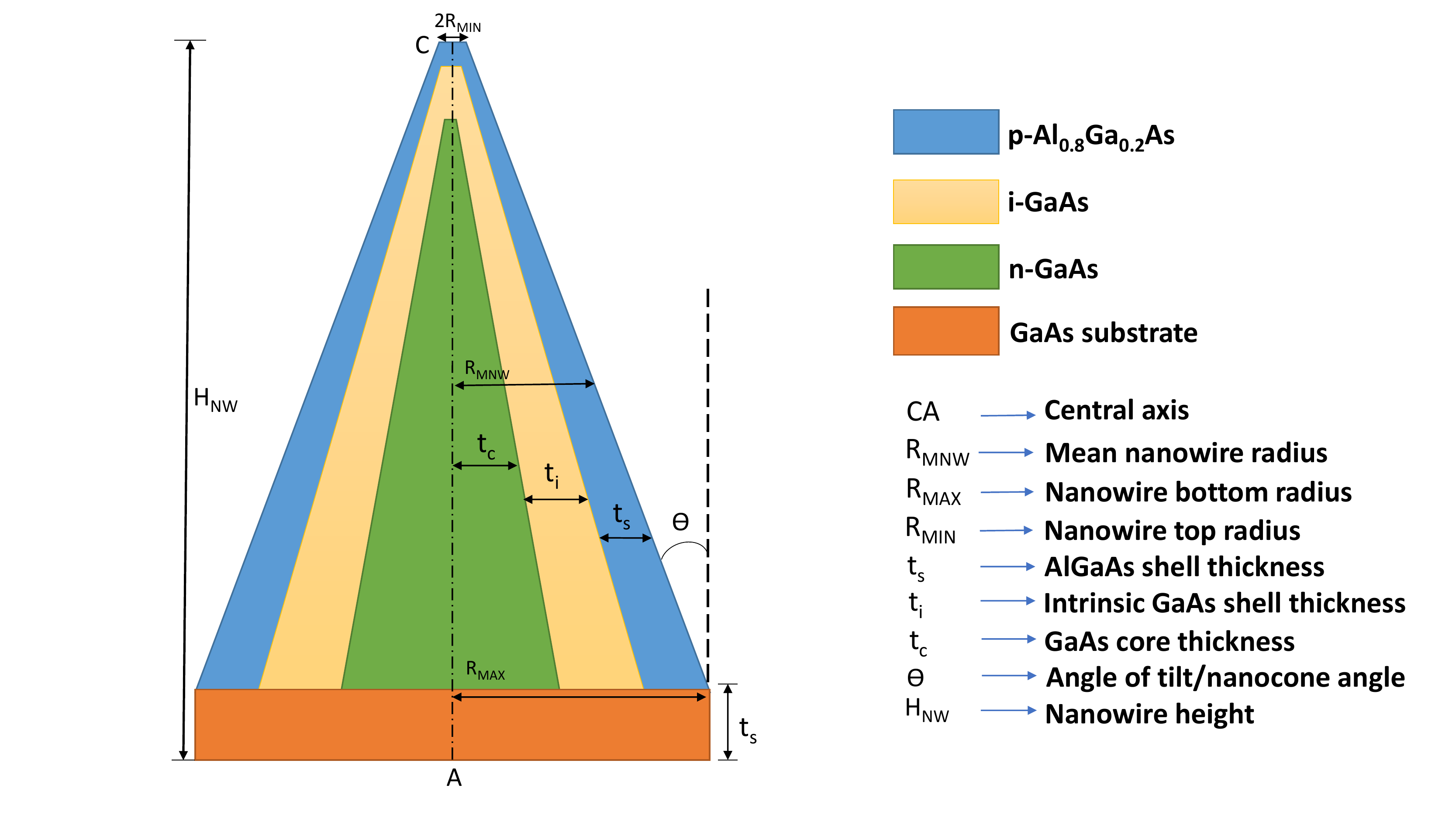}}
       
        \caption{(a) Perspective view (left) and top view (right) of the 3 x 3 nanocone array. P represents the period of the unit cell. (b) Cross-sectional schematic diagram of an unit cell nanowire that has been modelled for analysis. The dimensions are not to scale. }
        \label{fig:u}
\end{figure}



\begin{table}[t]
\centering
\caption{Simulation parameters for the nanocone solar cell}
\label{tab1}
\begin{tabularx}{3.5 in}{m{3 cm} m{2 cm} m{2 cm}}
\toprule
        Parameter  & Values (GaAs) & Values (AlGaAs) \\ 
\midrule
        Band Gap (eV)  & 1.42      & 2.07 \\ \\  \vspace{-0.15cm}
        Workfunction (eV)  & 4.78      & 4.563 \\ \\  \vspace{-0.15cm}
        DC Permittivity &  12.9 & 10.628\\ \\  \vspace{-0.15cm} 
        Radiative recombination coefficient ($C_{radiative}$) (cm$^{3}$.s$^{-1}$) & $7.2 \times 10^{-10}$ & $7.2 \times 10^{-10}$ \\ \\  \vspace{-0.15cm}
        SRH lifetimes ($\tau_{n}, \tau_{p}$) (ns)  & 1     & 1 \ \\
        Auger recombination coefficient ($A_{n}$, $A_{p}$) (cm$^{6}$.s$^{-1}$ )  & $1 \times 10^{-30}$        & $1 \times 10^{-30}$ \\ \\  \vspace{-0.15cm}
        Electron mobility ($\mu_{n}$) (cm$^{2}$.V$^{-1}$.s$^{-1}$ ) & $1200$  & $2300$      \\ \\  \vspace{-0.15cm}
        Hole mobility ($\mu_{p}$) (cm$^{2}$.V$^{-1}$.s$^{-1}$)   & $100$     & $146$ \ \\
        Electron relative effective mass & 0.067$m_{0}$ & 0.115$m_{0}$ \\ \\  \vspace{-0.15cm}
        Hole relative effective mass & 0.485$m_{0}$  & 0.598$m_{0}$\\ \\  \vspace{-0.15cm}
        Doping concentration (cm$^{-3}$) & $5 \times 10^{17}$  & $5 \times 10^{18}$   \\ \\  
    
\bottomrule \\
\end{tabularx}

\begin{tabularx}{3.5 in}{m{4 cm} m{4 cm} }
        Doping concentration (n-GaAs substrate) (cm$^{-3}$) & \tab \tab $5 \times 10^{17}$ \\ \\  \vspace{-0.15cm}
        Nanowire height ($H_{NW}$) ($\mu$m) & \tab \tab 2 \\ \\  \vspace{-0.15cm}
        Mean nanowire radius ($R_{MNW}$) (nm) & \tab \tab 180 \\ \\  \vspace{-0.15cm}
        Surface Recombination velocities ($S_{n}$, $S_{p}$) for `good passivation' ( cm.s$^{-1}$) & \tab  \tab 1300  \\ \\  \vspace{-0.15cm}
        Surface Recombination velocities ($S_{n}$, $S_{p}$) for `poor passivation' ( cm.s$^{-1}$)& \tab \tab  $1 \times 10^{7}$ \\ \\  \vspace{-0.15cm}
        Illumination &  \tab \tab  AM1.5 (1 sun)\\

\bottomrule
\end{tabularx}
\end{table}
\vspace{0 mm}
In this dissertation, we have modeled a pin GaAs/AlGaAs heterojunction nanocone solar cell (NCSC) array. For this purpose, a unit cell of the structure is first designed, and then periodic boundary conditions are used to simulate the entire square lattice \cite{mariani2013GaAs} . Fig~\ref{fig:u} shows the schematic diagram of the modeled nanocone solar cell. p-Al$_{0.8}$Ga$_{0.2}$As is used as the outer shell material; the inner regions are made of GaAs. The mean NC diameter is 360 nm with a period of 720 nm \cite{zhang2018photovoltaic}. For the optimal performance of cylindrical nanowires, the mean diameter should be 180 nm with D/P = 0.5 \cite{wen2011theoretical}. However, we won’t be able to appreciate the effect of nanocone angles for such small diameters. Moreover, due to tapering, the top diameter will be very small and will not be able to accept any wavelength modes. Therefore, a larger mean diameter is chosen. A period study is done on Section~\ref{sec:ps} to observe the effect of dense and sparse packing on nanowire performance. The height of the nanocones is taken to be 2 $\mu$m \cite{li2015influence, huang2012broadband}. The thickness of the GaAs substrate was made semi-infinite with appropriate PML boundary conditions \cite{li2015influence}. Doping is a very important parameter as it decides the behavior of the solar cell, the doping concentration of the p-type shell is $5 \times 10^{18}$ cm$^{-3}$  and the n-type core is  $5 \times 10^{17}$ cm$^{-3}$. While, it has been reported previously that high doping is favorable for core–shell NWSCs \cite{lapierre2011numerical, yao2014GaAs, li2015influence, wu2018optimization}, we have done systematic study \cite{li2015influence} to check if this doping satisfies the desired performance characteristics. For bulk GaAs substrate, concentration-dependent mobility is assumed. However, for the radial nanowires, due to the unavailability of such a systematic study that relates mobility to doping, the mobility is chosen from the work of Joyce $et$ $al.$ \cite{joyce2013electronic, li2015influence}. For AlGaAs, the mobility is chosen from \cite{shur1990physics}.
 The `nanocone angle' or the `angle of tilt' ($\theta$) is defined as the angle between the sidewall and the normal to the bottom surface. \\
 \tab For our work, four different models have been simulated; d1: Intrinsic region width = 10 nm, here the core and the shell have same radial thickness (t$_{c}$ = t$_{s}$, t$_{i}$ = 10 nm); d2: Intrinsic region, the shell and the core have the same thickness (t$_{c}$ = t$_{s}$  = t$_{i}$);  d3: The thickness of the intrinsic region is double to that of the  and shell (2t$_{c}$ = 2t$_{s}$  = t$_{i}$); d4: The thickness of the intrinsic region is triple to that of the core and shell (3t$_{c}$ = 3t$_{s}$  = t$_{i}$). For these designs, as we go from design `d1' to design `d4', the volume of AlGaAs decreases and volume of GaAs increases, also the width of the i-GaAs shell increases, which will result in more absorption; however, if the photogenerated carriers can be efficiently extracted or not, is what we wish to observe. For each, we have changed the angle from 0 to 9$^{\circ}$ by varying the top and bottom radius ( R$_{MIN}$ and R$_{MAX}$), respectively keeping the mean nanowire radius (R$_{MNW}$) constant.\\
 \tab  For optical simulations, `Lumerical FDTD' software package is used. The absorption per unit volume is calculated using the Poynting vector $\vec{P}$ as:
 \begin{equation}
P a b s=-0.5 \operatorname{real}(\vec{\nabla} \cdot \vec{P})
\end{equation}
Which can be written in a more numerically stable form : 
\begin{equation}
P a b s=0.5 \operatorname{real}\left(i \omega \vec{E} \cdot \vec{D}\right) = 0.5 \omega \varepsilon^{\prime}|\vec{E}|^{2}
\end{equation}
Where, $\varepsilon^{\prime}$ is the imaginary part of permittivity, $\omega$ is the angular frequency of the incident light, and $E$ is the electric field intensity. Assuming that each photon absorbed generates an equal number of electron-hole pairs, we can write the photogeneration rate as: 
\begin{equation}
G_{p h}=\frac{|\vec{\nabla} \cdot \vec{S}|}{2 \hbar \omega}=\frac{\varepsilon^{ \prime}|\vec{E}|^{2}}{2 \hbar}
\end{equation}
Where $\hbar$ is the reduced Plank's constant, $ G_{p h}$ is weighed by the AM 1.5G solar spectrum and integrated over the entire simulation spectrum. The complex refractive index of GaAs and AlGaAs has been taken from the Sopra database. The reflectance and transmittance spectra normalized to the source power are measured using the `Frequency-Domain Field and Power' monitors at the top and bottom of the simulation region respectively.\\
\tab For electrical simulations, the software package `Lumerical CHARGE' is used. The simulated photogeneration profile is incorporated into a finite-element mesh of the nanocones, which self-consistently solves the carrier continuity equations coupled with the nonlinear Poisson's equations in 3D. The details and working of the FDTD and CHARGE solver can be found here \cite{charge,fdtd}. For our study, we have assumed the interface between GaAs and AlGaAs is perfect without any additional recombination centers, which can be achieved by lattice-matched epitaxy \cite{molenkamp1988very}. The surface effects have only been considered for the interfaces between air and the NW. For AlGaAs as the outer surface, a good SRV of 1300 cm.s$^{-1}$ is chosen. For GaAs surfaces, two SRV cases are considered; good SRV: 1300 cm.s$^{-1}$ and poor SRV: $10^{7}$ cm.s$^{-1}$ with the same SRV for both electrons and holes. Radiative, Auger, and SRH recombination models and bandgap narrowing models are also considered. We have used top and bottom ohmic contacts for our structure \cite{lapierre2011numerical}. The auger recombination coefficients, SRH recombination lifetime, and radiative recombination rates are taken to be the same for GaAs and AlGaAs \cite{wu2018optimization, lapierre2011numerical, li2015influence}. The various key simulation parameters are listed in Table~\ref{tab1}.


\section{Analysis and Results}
\label{sec:ar}


\subsection{Nanocone Angle Study}
\label{sec:nas}
For this study, the design model `d1' is chosen. In this design, the intrinsic region has a constant thickness of 10 nm, the thickness of core and shell is the same throughout the nanocone's height. The nanocone angle is varied from 0 to 9 degrees by varying the top and bottom radius, keeping the mean radius constant at 180 nm with a D/P (duty cycle) = 0.5 \cite{zhang2018photovoltaic}.


\subsubsection{Optical Properties}
\begin{figure}[t]
  \centering
  \includegraphics[width=3.5 in]{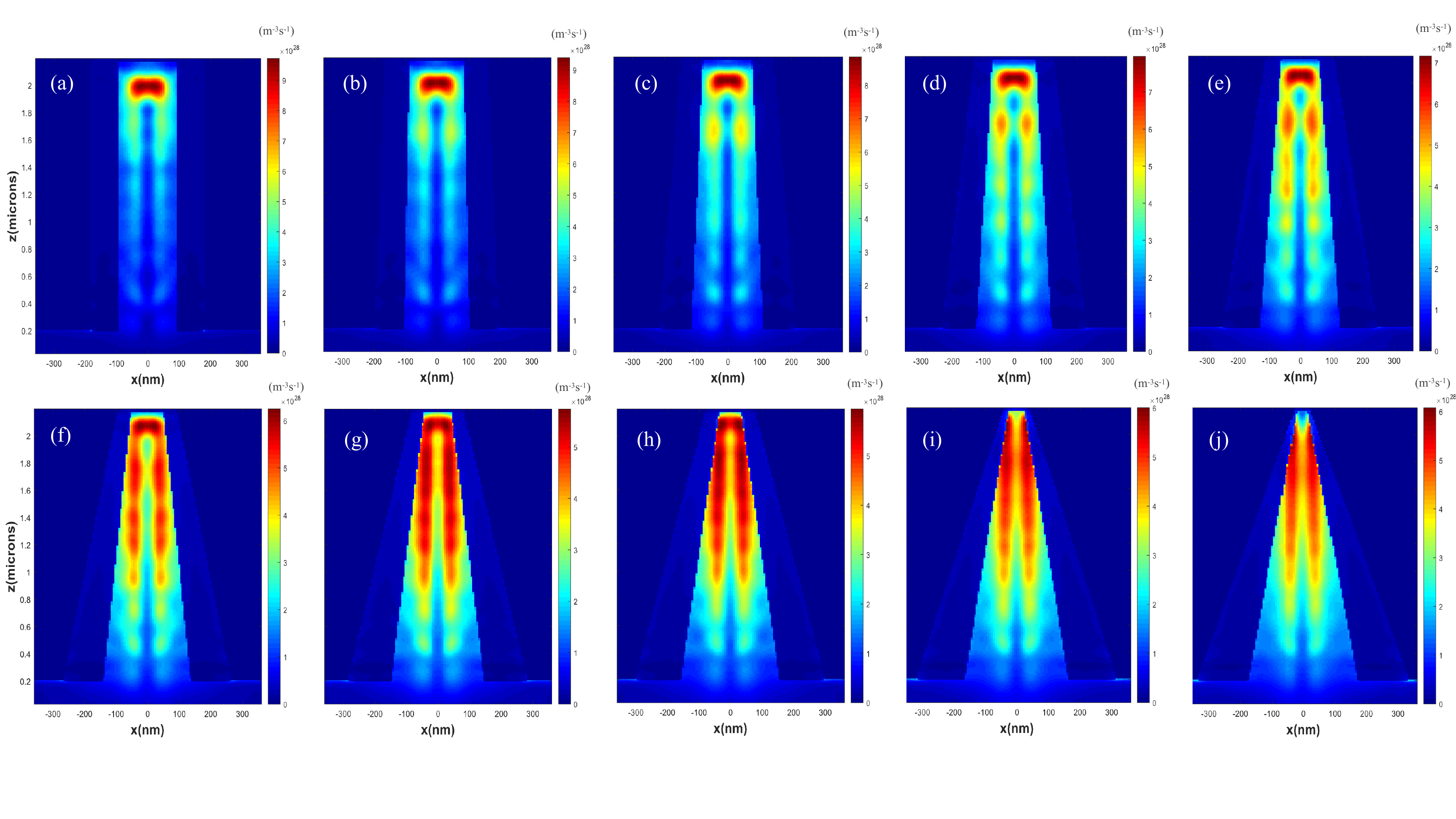}\\
  \caption{The photogeneration profiles of radial pin nanocone design `d1' for (a) $\theta$ = 0$^{\circ}$, (b) $\theta$ = 1$^{\circ}$, (c) $\theta$ = 2$^{\circ}$, (d) $\theta$ = 3$^{\circ}$, (e) $\theta$ = 4$^{\circ}$, (f) $\theta$ = 5$^{\circ}$, (g) $\theta$ = 6$^{\circ}$, (h) $\theta$ = 7$^{\circ}$, (i) $\theta$ = 8$^{\circ}$, (j) $\theta$ = 9$^{\circ}$.}
  \label{fig:lo}
\end{figure}
\begin{figure}[t]
  \centering
  \includegraphics[width=3.5 in]{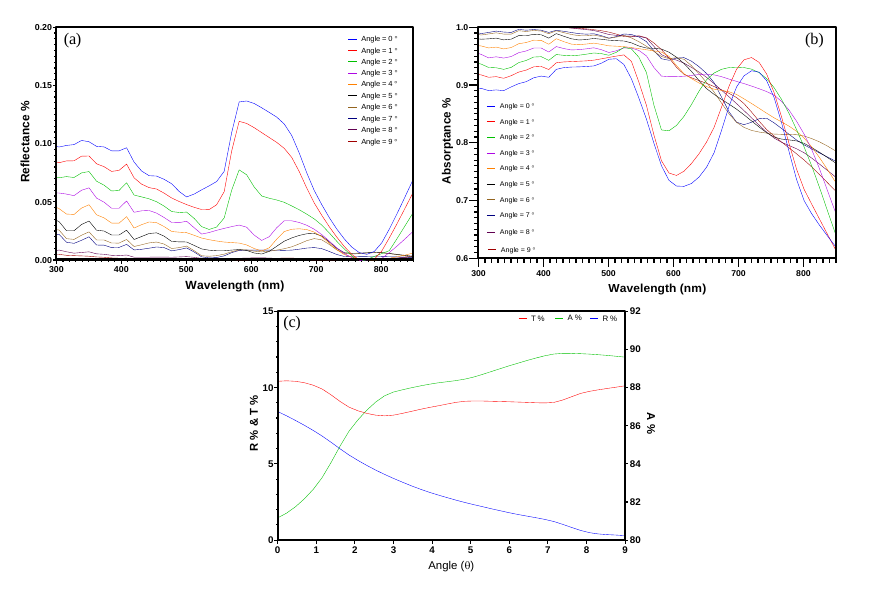}\\
  \caption{(a) Reflectance, (b) Absorptance, (c) The integral of absorptance, reflectance and transmittance of the radial pin junction nanocone `d1' for different nanocone angles.}
  \label{fig:la}
\end{figure}
Fig~\ref{fig:lo} shows the photogeneration profiles of the nanocone solar cell for different nanocone angles. It is observed that when the angle of the nanocone is less, the photogeneration hotspot is mostly confined to the top of the nanowires. By introducing a tapered structure, we observe that as the nanocone angle increases, the photon absorption shifts downward and spreads across the nanocone's length, leading to enhanced effective absorption. However, this process increases the absorption till a certain angle, and after a certain angle, the absorption tends to decline. From the nanocone's reflectance and absorptance spectra (ref. Fig~\ref{fig:la}), we can see that the reflectance decreases throughout the entire wavelength range as the nanocone angle increases, whereas the absorption increases for lower wavelengths modes ($\sim$ 300 - 700 nm). However, for large angles, the absorptance tends to decrease at the large wavelength regime ($\geq$ 700 nm). The NW arrays' anti-reflection ability can be attributed to the low filling ratio, which reduces the effective refractive index and offers a good impedance match between AlGaAs and air \cite{zhang2018photovoltaic, wu2017efficient}. For the nanocone arrays with a large slope angle, the filling ratio at the top of the array is extremely low, leading to a nearly perfect impedance match with air and almost zero reflection. Light absorption in NWs is dominated by resonant modes, which are closely related to the NW diameter \cite{zhang2018photovoltaic, anttu2010coupling, zhan2014enhanced}. In nanocones, the diameter of the structure continually changes across the nanowire height, with the top diameter being significantly smaller than the bottom diameter, and due to this unique geometry, it can support only a few long-wavelength modes in the top and absorption (particularly for long wavelengths) shifts towards to the thicker middle of the structure \cite{zhang2018photovoltaic}. Thus there is an even distribution of light absorption throughout the structure. This increase in absorption with the increase in nanocone angle is supported by a decrease in the reflectance attributed to different RI mismatches at the air/AlGaAs interface and enhanced multiple scattering effects.
For large angles, the nanocone top becomes too thin, and the reduction in absorption in the top regions is more dominant than the gain in the middle area, thereby resulting in a decrease in absorptance in the long-wavelength regime. As the absorption shifts downward, the effective absorption length increases, and as the i-region exists in the radial direction, this downward shift leads to an enhanced overlap between the i-region and photogeneration. Therefore, the effective absorption is also believed to be increased coupled with an improvement in carrier extraction.
\begin{figure}[t]
  \centering
  \includegraphics[width=3.4 in]{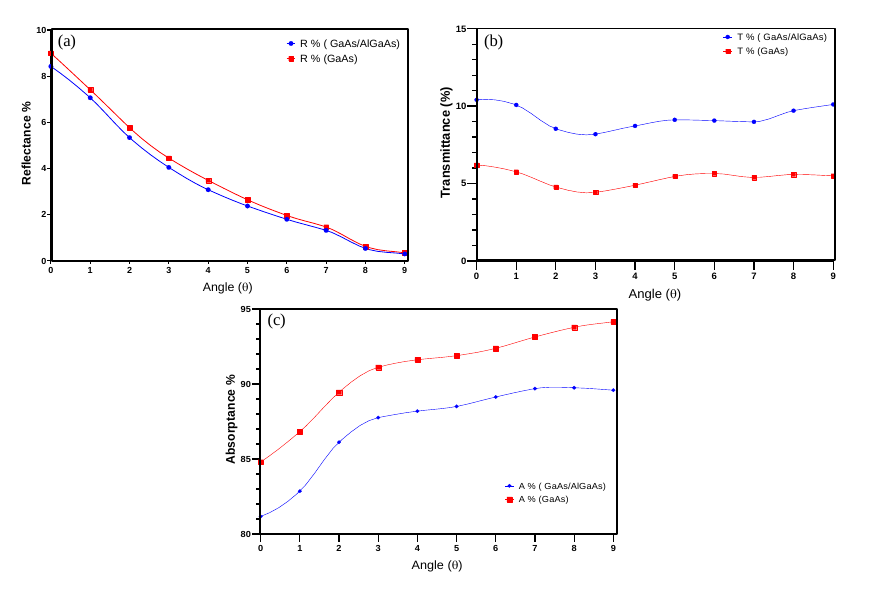}\\
  \caption{The integral of (a) reflectance, (b) transmittance, (c) absorptance of the nanocone `d1' with GaAs shell and AlGaAs shell for different nanocone angles.}
  \label{fig:lc}
\end{figure}
\tab Fig~\ref{fig:lc} compares the properties of GaAs/AlGaAs nanocones with their GaAs counterparts, we can see that GaAs show improved absorption and lower transmittance for all the nanocone angles,  attributed to its larger absorptivity than AlGaAs. However, this does not lead to an improvement in its photovoltaic performance.


\subsubsection{Electrical Properties}
\begin{figure}[b]
  \centering
  \includegraphics[width=3.5 in]{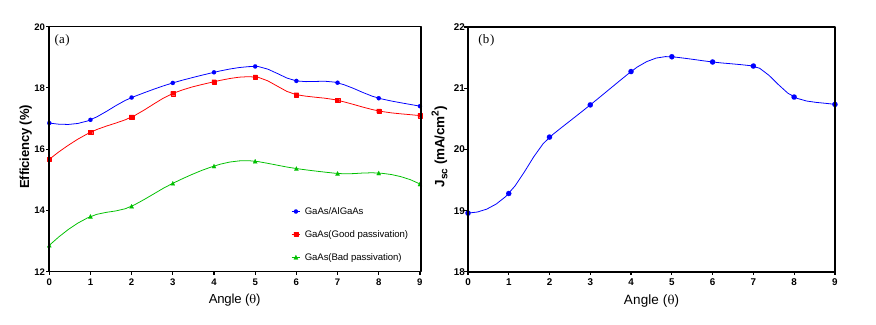}\\
  \caption{(a) Conversion efficiency of the radial pin junction nanocone `d1' with AlGaAs shell, GaAs shell (Good and Bad passivation) for different nanocone angles. (b) Short circuit current density (J$_{sc}$) of the radial pin junction nanocones with AlGaAs shell.}
  \label{fig:ld}
\end{figure}

In Fig~\ref{fig:ld}-a, we observe the efficiency of the NCSC varies vs the nanocone angle for structures with AlGaAs shell with good passivation, GaAs shell with good passivation, and GaAs shell with poor passivation. It can be seen that the designs with AlGaAs shells have superior performance than the GaAs counterparts for the entire range of angles. For small angles, the optical generation hotspot is situated in the top of the solar cell \cite{zhang2018photovoltaic, wu2018optimization} but due to lack of electric field to drive those photogenerated carriers, the carrier extraction will be poor \cite{anttu2010coupling, fossum1982physical, fossum1983carrier}. This results in significant recombination of photocarriers and loss of absorbed power. This loss is less when AlGaAs are used as the outer shell material. Due to its large bandgap and less absorptivity, minimal absorption will occur in AlGaAs, and most of the absorption will stay confined to the inner GaAs regions \cite{wu2018optimization}. Thus, reducing the surface recombination losses present in GaAs and improving the solar cell performance \cite{tajik2011sulfur, mariani2013GaAs, aaberg2015GaAs, chang2012electrical}. AlGaAs also acts as a barrier preventing the photogeneration carriers that are generated in the inner GaAs regions from reaching the surface, and
it also acts as an effective passivation layer reducing surface recombination \cite{chang2012electrical, demichel2010impact}.\\

\tab As the nanocone angle increases, the photogeneration is more evenly spread out throughout the length of the structure and therefore it overlaps with the radial intrinsic region, thereby allowing efficient carrier extraction. As a result, the $J_{sc}$ increases with the increase in angle from 0 - 5 degrees (Fig~\ref{fig:ld}-(b)) with a corresponding improvement in conversion efficiency. However, after a certain angle, the efficiency reaches its maximum value, and then it starts to decrease, which is probably because the useful photogeneration reduces. At large angles, the top of the structure becomes very thin to support long wavelength modes, and even though the longer modes get absorbed towards the middle, it is offset by the decrease in absorption in the top, thus resulting in a reduction of efficiency.


\subsection{Nanocone Design Study}
\label{sec:nds}

Since the overlap of the optical absorption with the i-GaAs region is the key to the nanocone's performance, four designs have been tested to find the optimal thickness profile of the i-GaAs shell. In design 1, we have a constant intrinsic region. From designs 2 - 4, the intrinsic GaAs region thickness increases across the nanocone's height. This is done to obtain the profile of the intrinsic region that provides a more efficient overlap of the photogeneration with the intrinsic region and better optical and electrical properties.

\subsubsection{Optical Properties}
\label{sec:op}
\begin{figure}[t]
  \centering
  \includegraphics[width=3.5 in]{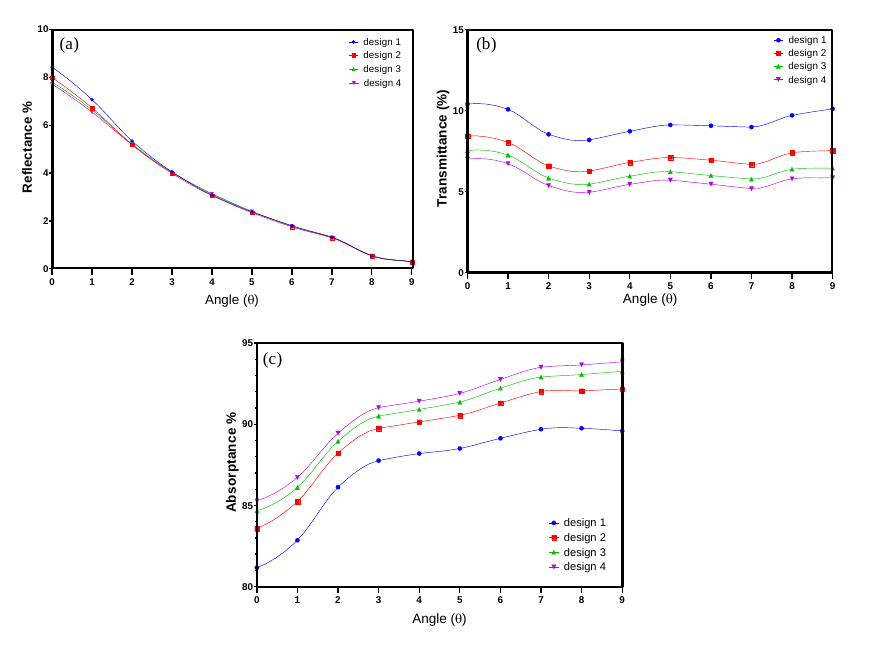}\\
  \caption{The integral of (a) reflectance, (b) transmittance, (c) absorptance of the nanocone with different nanocone angles for different design configurations.}
  \label{fig:l}
\end{figure}
On Fig~\ref{fig:l}, the effect of different designs on the optical properties (the net absorptance, reflectance, and transmittance) of the NCSCs are observed. It can be seen that the design `d1' has the lowest absorption and highest transmittance for all the nanocone angles, followed by `d2' with better absorption and lower transmission, and with `d4' having the best absorption and least transmission. Such a trend is expected as `d1' has the highest volume of AlGaAs and `d4' has the lowest, and GaAs have much larger light-absorbing properties than AlGaAs. The structure's reflectance for different configurations remains almost the same because the overall geometry is not changed.
Therefore, in terms of the optical properties, design `d4' has a better performance than the other designs.

\subsubsection{Electrical Properties}
\label{sec:ep}

\begin{figure}[t]
  \centering
  \includegraphics[width=3.5 in]{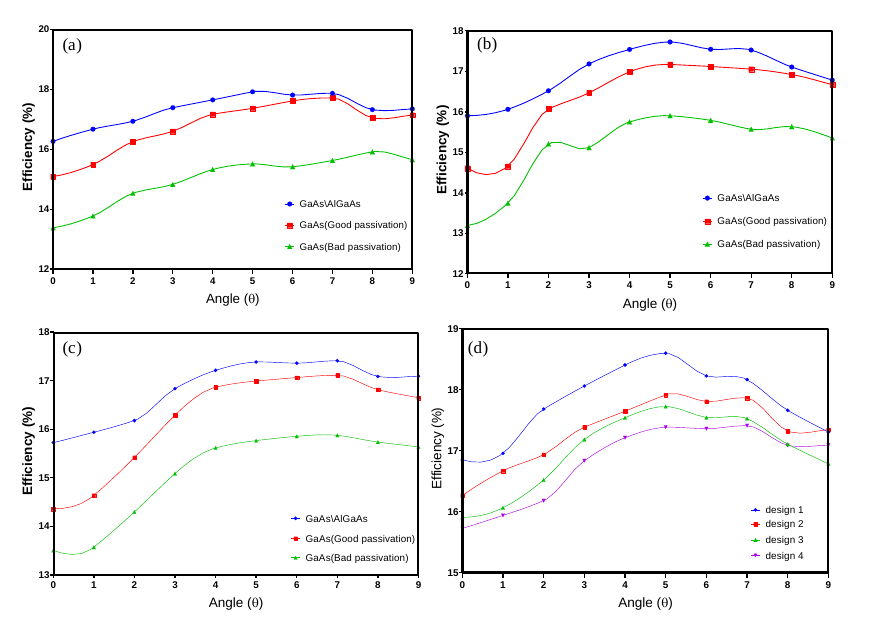}\\
  \caption{Conversion efficiency of the radial pin junction nanocone design (a) `d2', (b)`d3', (c) `d4' with AlGaAs shell, GaAs shell (Good and Bad passivation) for different nanocone angles. (d) Conversion efficiency of the radial pin junction nanocone (AlGaAs shell) with different nanocone angles for different design configurations. }
  \label{fig:zz}
\end{figure}
Fig~\ref{fig:zz}-a,b,c shows the conversion efficiencies of different NCSC arrays for designs `d2', `d3' and `d4'. We can see that the different designs follow a similar profile; initially, the efficiency increases, then it gradually starts to decrease (similar to the design `d1'). Further, it's observed that NCs with AlGaAs shell has better efficiency than the one with a GaAs shell for all the cases. Hereby emphasizing the improved performance of GaAs/AlGaAs nanocones. We can apply the same reasoning as in the case of `d1'. In Fig~\ref{fig:zz}-d, the conversion efficiencies for all the designs with AlGaAs shell have been compared. In contrast to the trend in the optical properties, we see that the overall efficiency of design `d1' is the greatest for all the angles, and efficiency decreases as we go from design `d1' to `d4'. The efficiency increases monotonically from 0$^{\circ}$ - 5$^{\circ}$ then it saturates followed by a decrease for larger angles.

\begin{figure}[t]
        \centering
        \subfloat[]{\label{fig:fa}\includegraphics[width=3.45in]{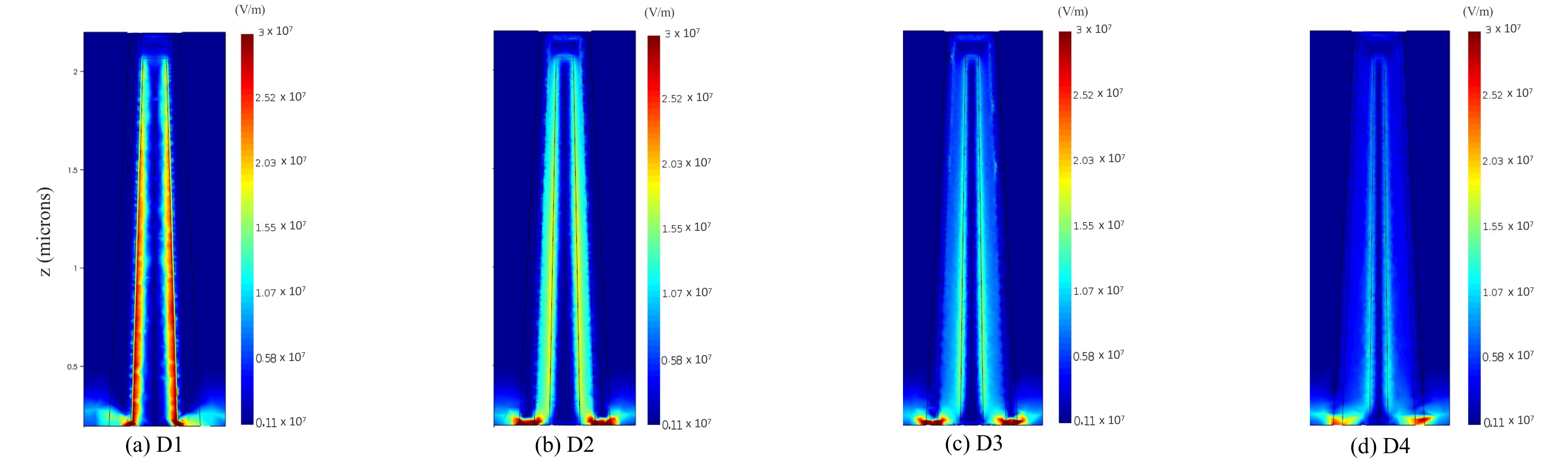}}
        \qquad
        \subfloat[]{\label{fig:fb}\includegraphics[width=3.38in]{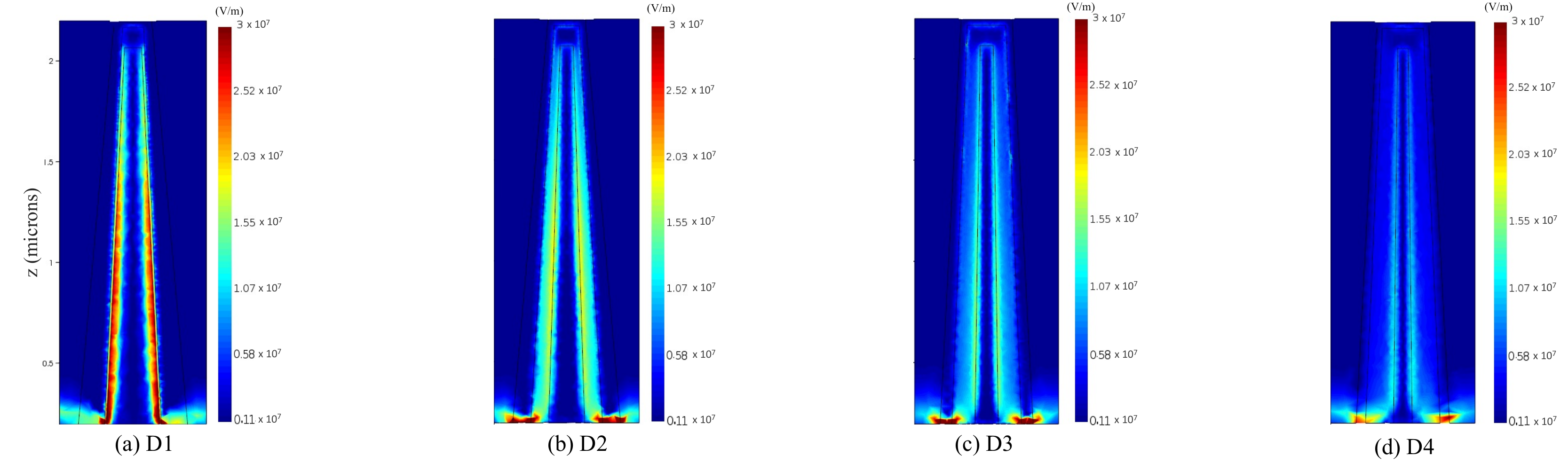}}
        \qquad
        \subfloat[]{\label{fig:fc}\includegraphics[width=3.45in]{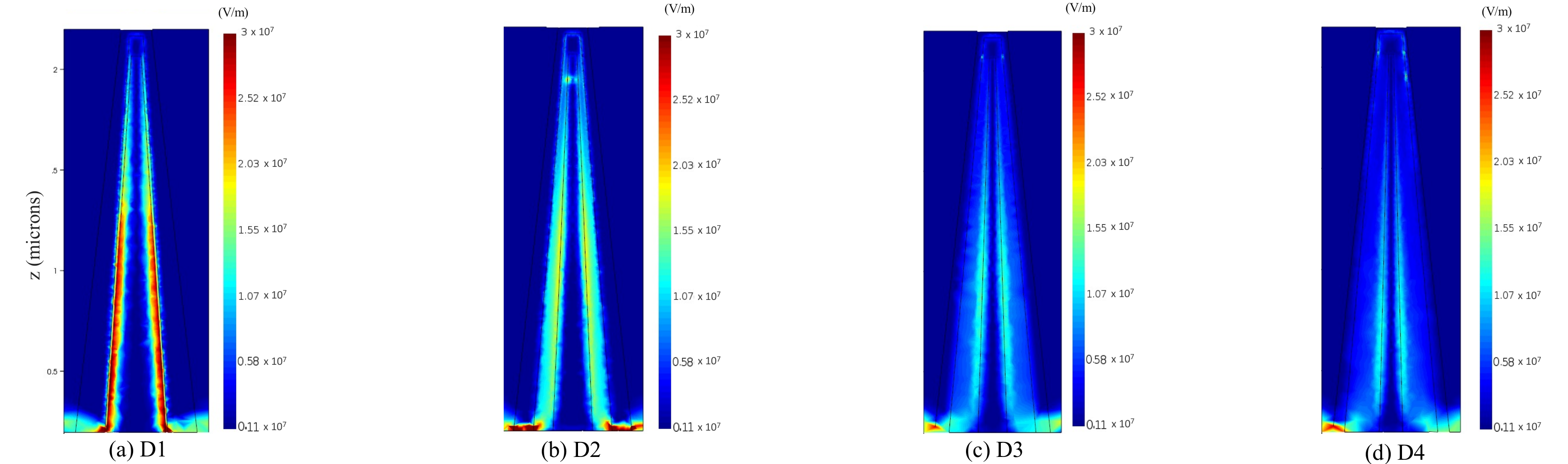}}
        \caption{Cross-sectional view of the magnitude of the electric field of the four pin junction nanocone designs for nanocone angles = (a) 3$^{\circ}$, (b) 5$^{\circ}$, (c) 9$^{\circ}$.  }
        \label{fig:f}
\end{figure}
\tab One reason for such a contrasting behavior is the presence of the built-in electric field. We can see it from Fig~\ref{fig:f}- a,b,c that the electric field mostly exists around the interface between the (n-core) - (i-shell) interface but as the thickness of the i-GaAs region increases from design `d1' – `d4' (for nanocone angles: 3$^{\circ}$, 5$^{\circ}$, 7$^{\circ}$), the magnitude of the electric field continually decreases and has a maximum value for the case of `d1' (constant intrinsic region). As a result, when we go from `d1' to `d4', the extraction of the photogenerated carriers decreases, and the recombination rate rises. Further, the built-in electric field generated in the 1st case (`d1') may overlap with the maximal photogeneration region. Compared to the other designs, a large shell allows the photogeneration to happen dominantly in the inner GaAs regions close to the interface where the electric field exists and it also pushes the surface away thus reducing recombination. Both these factors lead to improved performance of `d1', and thus having a constant i-shell thickness can give substantially better results than a tapered radially varying i-shell.


\subsubsection{Design Optimization}
\label{sec:do}
\begin{figure}[t]
  \centering
  \includegraphics[width=3.5 in]{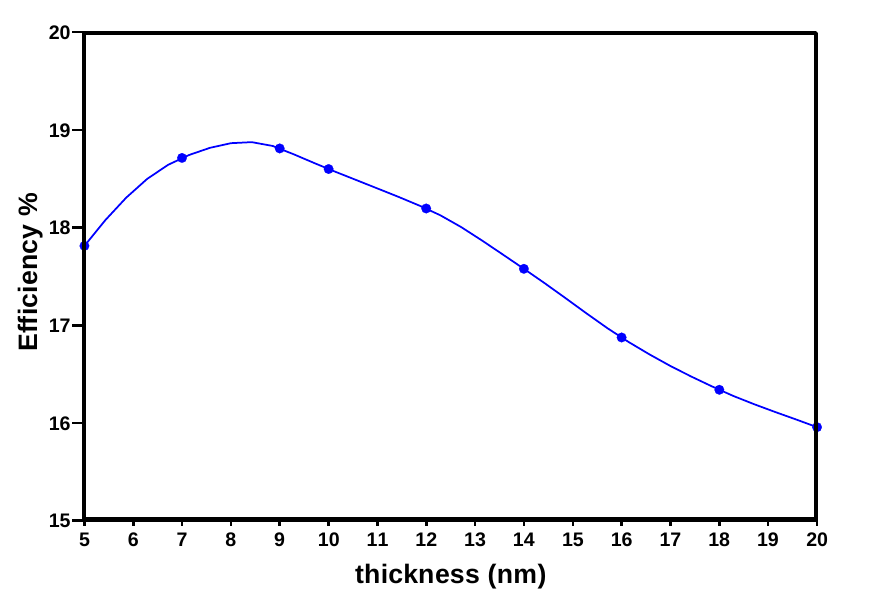}\\
  \caption{Conversion efficiency of the radial pin junction nanocone `d1' with different i-GaAs shell thickness.}
  \label{fig:z}
\end{figure}
Now that we have decided on the optimal design configuration, we now look forward to optimizing it. Since the key to this design is the constant thickness of the intrinsic shell, we need to find the most optimal thickness which can give us the best performance. For that, we have done a parametric sweep of the thickness of the i-shell (t$_{i}$) from 5 nm to 20 nm and observed its effect on the solar cell efficiency. From Fig~\ref{fig:z}, we can see that, with the increase in t$_{i}$, the efficiency first increases and attains the maximum at a thickness of 8 - 9 nm and then decreases.
\begin{figure}[t]
  \centering
  \includegraphics[width=3.5 in]{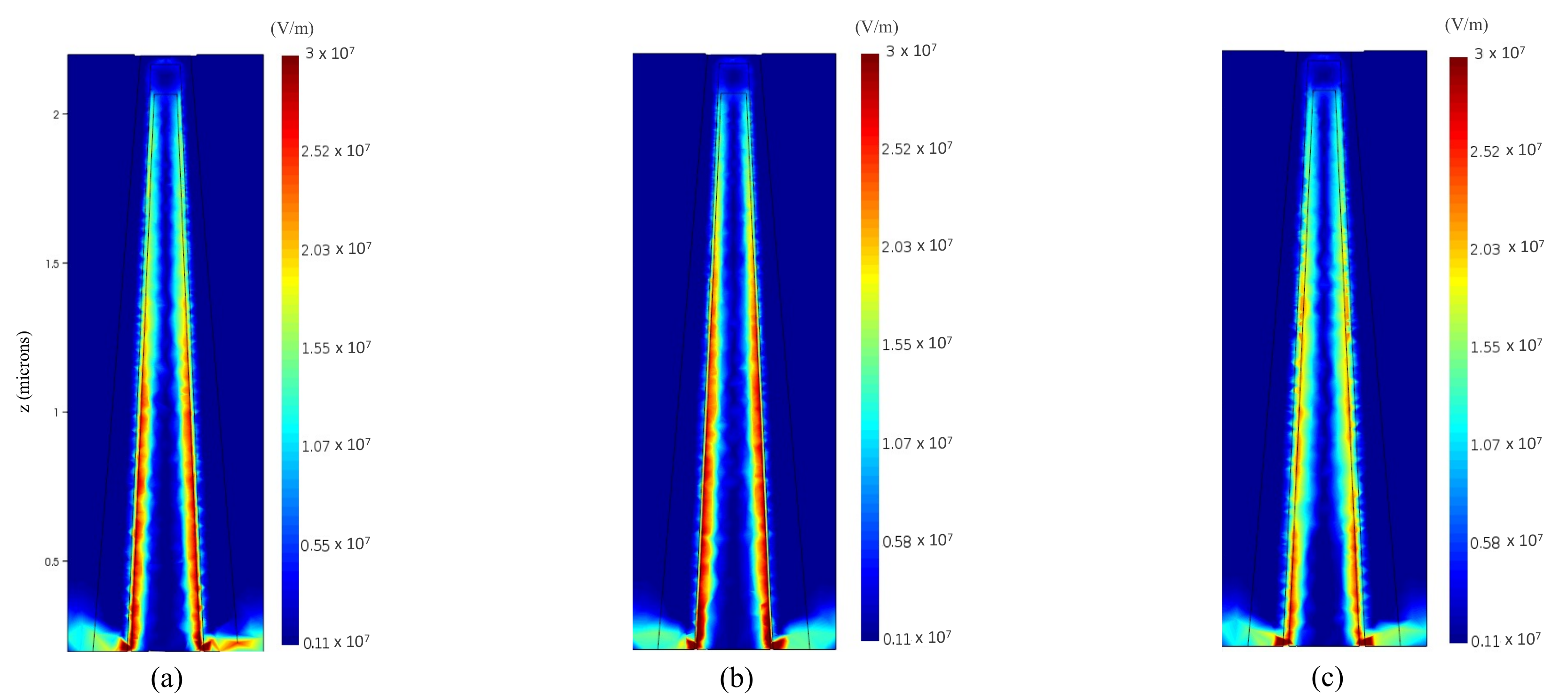}\\
  \caption{Cross-sectional view of the magnitude of the electric field of the radial pin junction nanocone `d1' with i-GaAs shell thickness: (a) 5 nm, (b) 9 nm, (c) 20 nm.}
  \label{fig:x}
\end{figure}
The possible reason for this type of efficiency profile could be due of a trade-off between photogeneration and carrier extraction. As the thickness of i-GaAs increases, the thickness of the AlGaAs shell decreases, and thus the overall absorption (which primarily happens in GaAs) increases leading to improved performance. On the other hand, with a large increase in thickness, the electric field magnitude across the pin junction decreases as seen from Fig~\ref{fig:x}, leading to poor carrier extraction and degrading efficiency. Also, one crucial factor is the overlap of the built-in electric field with photogeneration hotspots, which may also change depending on the position and thickness of the radial junction. Therefore, considering all these constraints, we get an optimized thickness satisfying all the factors.

\subsection{Period Study}
\label{sec:ps}
\begin{figure}[b]
  \centering
  \includegraphics[width=3.6 in]{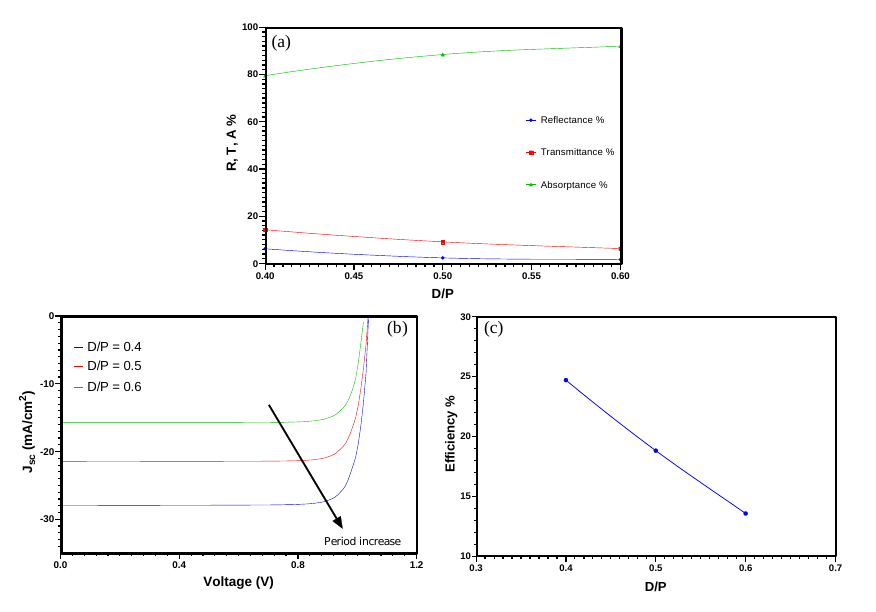}

  \caption{(a) The integral of absorptance, reflectance and transmittance, (b) IV curves  (c) Conversion efficiency of the radial pin junction nanocone `d1' for different diameter/period ratios keeping the mean diameter constant = 180 nm.}
  \label{fig:a}
\end{figure}


In this section, we observe the effect of dense and sparse packing of the nanocone array on its optical and electrical properties. Fig~\ref{fig:a}-a shows the net absorptance, reflectance, and transmittance of the nanocone array for different unit cell periods, keeping the cell diameter constant at 180 nm. It is observed that as the period decreases (D/P increases), there is an increase in absorption with an overall absorptance of $\sim$ 80 $\%$ at D/P = 0.4 to a $\sim$ 90 $\%$ at D/P = 0.6 with a corresponding decrease in total reflection and transmission. The difference in the anti-reflective properties between the dense and sparse models is attributed to the different effective refractive index mismatches at the air/SC interface \cite{jung2010strong} and enhanced multiple scattering effects in the dense NCs \cite{chattopadhyay2010anti, bai2014one}. \\
\tab From this observation alone, one can easily expect the nanowires having smaller periods to perform better. In order to verify that, we need to analyze the structure's photovoltaic properties as well. Fig~\ref{fig:a}-b shows the IV curves and the conversion efficiencies of the GaAs/AlGaAs nanocone solar cell arrays as D/P is varied from 0.6 to 0.4 with a constant diameter of 180 nm. 
In contrast to the optical properties, we see a huge increase in the efficiency from around 13 $\%$ to 27 $\%$ as the period is increased, i.e. for sparse NCSCs, the higher efficiency is obtained despite the better anti-reflective properties of the dense NCs. This change in efficiency comes from the improvement in $J_{sc}$ as seen from Fig~\ref{fig:a}-c. This behavior is similar to the behavior of SiNW/ PEDOT: PSS cells \cite{bai2014one}. With the same reasoning, we can say that the increased $J_{sc}$ can be attributed to the more effective filling of AlGaAs into the sparse NCs versus the dense NCSCs. This improvement in efficiency results from an improved optical generation rate and the corresponding decrease in the net recombination rate of the photogenerated carriers in the sparse structures.
\begin{figure}[t]
        \centering
        \includegraphics[width=3.5in]{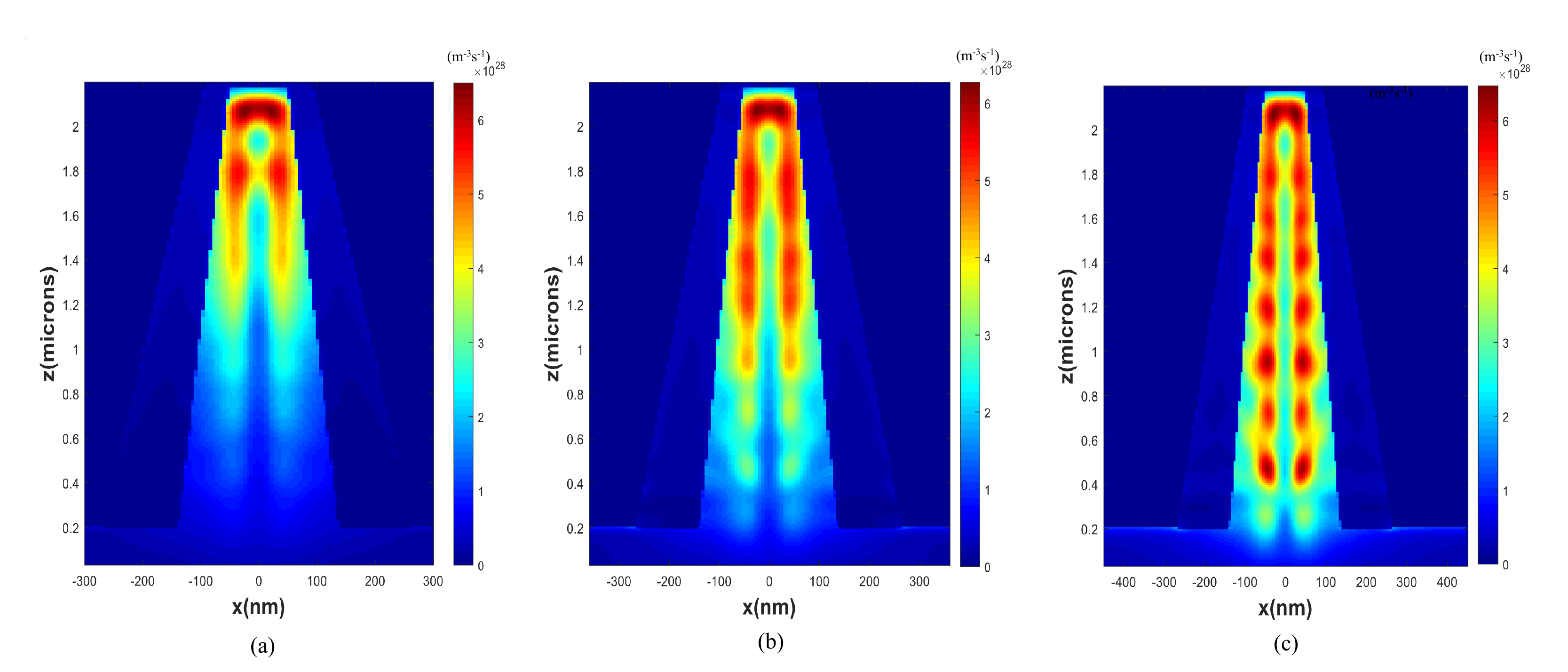}
        \caption{Photogeneration profiles of the radial pin junction nanocone `d1' with D/P : (a) 0.6, (b) 0.5, (c) 0.4.}
        \label{fig:c}
         \vspace{-0 cm}
\end{figure}
\begin{figure}[t]
  \centering
  \includegraphics[width=3.5 in]{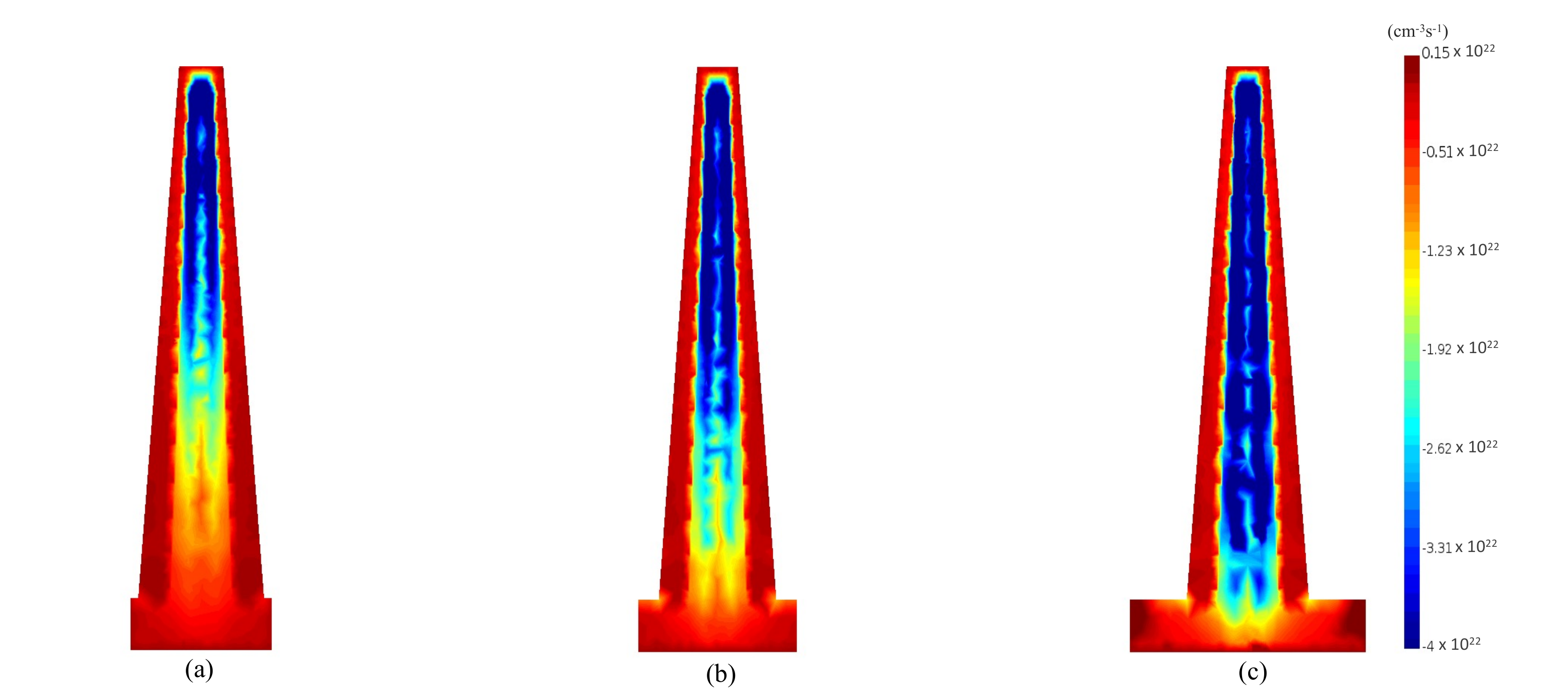}\\
  \caption{Net photo-carrier recombination profiles of the radial pin junction nanocone `d1' with D/P : (a) 0.6, (b) 0.5, (d) 0.4.}
  \label{fig:d}
  \vspace{-0 cm}
\end{figure}
For sparse structures, even though the overall absorption is less; the photogeneration profile, which comes from the useful absorption spectra, spreads across the entire structure and results in improved effective absorption and photogeneration as seen from Fig~\ref{fig:c}. Here, we can see that when the D/P ratio is low (the period is large), the photogeneration spreads out across the entire length of the solar cell. This helps us attain a longer absorption length. Which in this case, is better than the photogeneration at large nanocone angles, as it's obtained without compensating for the loss of absorption due to the thinner top region. This improved photogeneration profile coupled with its strong overlap with the pin junction's electric field leads to a decrease in net photogenerated carrier recombination as seen from Fig~\ref{fig:d}. These coupled effects lead to an improvement in the short circuit current ($J_{sc}$) of the solar cell, which is reflected in its efficiency. 


\section{Conclusions}
\label{sec:conc}
An extensive study of the optical and photovoltaic properties of pin junction GaAs/AlGaAs nanocone array solar cells has been done. It is observed that with the increase in the nanocone angle, the photogeneration, which is generally confined to the top for cylindrical NWs, spreads out along the length
of the structure. This happens because the top of the cell becomes too thin to support many long-wavelength modes, which get absorbed towards the bottom, increasing the absorption length. However, having a very large angle is also detrimental to the cause; as for angles $\geq$ 5 - 7 degrees, the loss of light absorption at the top becomes greater than the light absorbed at the bottom. Four different designs have been studied to observe the effect of the variation of the intrinsic shell position and thickness on the nanowire performance. The efficiency curves for all the structures follow a similar trend, i.e. they increase for small angles, then saturate and decrease for larger angles. It is observed that the design with a constant i-region delivers the best electrical results because of its strong built-in electric field throughout the structure and its overlap with the photogeneration. For all the cases, the NCs with AlGaAs shells have better overall performance than NCs with GaAs shells, highlighting the importance of AlGaAs passivation and confinement.  The thickness of the i-region is further optimized to find the configuration for best solar cell performance, and it is seen that a thickness of 8 - 9 nm provides the best efficiency. A period study is done to see the effect of dense and sparse packing of NCs in the array, and it has been observed that sparse NCs with larger periods had much larger efficiency than the densely packed ones because of more effective filling of AlGaAs, large effective photogeneration throughout the structure (it behaves similar to the cells having large nanocone angles w/o the loss incurred due to the thin top region) and the decreased net recombination of the photogenerated carriers.



\bibliography{references}

\begin{thebibliography}{10}
\providecommand{\url}[1]{#1}
\csname url@samestyle\endcsname
\providecommand{\newblock}{\relax}
\providecommand{\bibinfo}[2]{#2}
\providecommand{\BIBentrySTDinterwordspacing}{\spaceskip=0pt\relax}
\providecommand{\BIBentryALTinterwordstretchfactor}{4}
\providecommand{\BIBentryALTinterwordspacing}{\spaceskip=\fontdimen2\font plus
\BIBentryALTinterwordstretchfactor\fontdimen3\font minus
  \fontdimen4\font\relax}
\providecommand{\BIBforeignlanguage}[2]{{%
\expandafter\ifx\csname l@#1\endcsname\relax
\typeout{** WARNING: IEEEtran.bst: No hyphenation pattern has been}%
\typeout{** loaded for the language `#1'. Using the pattern for}%
\typeout{** the default language instead.}%
\else
\language=\csname l@#1\endcsname
\fi
#2}}
\providecommand{\BIBdecl}{\relax}
\BIBdecl

\bibitem{guo2010catalyst}
W.~Guo, M.~Zhang, A.~Banerjee, and P.~Bhattacharya, ``Catalyst-free ingan/gan
  nanowire light emitting diodes grown on (001) silicon by molecular beam
  epitaxy,'' \emph{Nano letters}, vol.~10, no.~9, pp. 3355--3359, 2010.

\bibitem{qian2005core}
F.~Qian, S.~Gradecak, Y.~Li, C.-Y. Wen, and C.~M. Lieber, ``Core/multishell
  nanowire heterostructures as multicolor, high-efficiency light-emitting
  diodes,'' \emph{Nano letters}, vol.~5, no.~11, pp. 2287--2291, 2005.

\bibitem{duan2003single}
X.~Duan, Y.~Huang, R.~Agarwal, and C.~M. Lieber, ``Single-nanowire electrically
  driven lasers,'' \emph{Nature}, vol. 421, no. 6920, pp. 241--245, 2003.

\bibitem{saxena2013optically}
D.~Saxena, S.~Mokkapati, P.~Parkinson, N.~Jiang, Q.~Gao, H.~H. Tan, and
  C.~Jagadish, ``Optically pumped room-temperature gaas nanowire lasers,''
  \emph{Nature photonics}, vol.~7, no.~12, pp. 963--968, 2013.

\bibitem{wilhelm2017broadband}
C.~E. Wilhelm, M.~I.~B. Utama, G.~Lehoucq, Q.~Xiong, C.~Soci, D.~Dolfi,
  A.~De~Rossi, and S.~Combri{\'e}, ``Broadband tunable hybrid photonic
  crystal-nanowire light emitter,'' \emph{IEEE Journal of Selected Topics in
  Quantum Electronics}, vol.~23, no.~5, pp. 1--8, 2017.

\bibitem{dai2014GaAs}
X.~Dai, S.~Zhang, Z.~Wang, G.~Adamo, H.~Liu, Y.~Huang, C.~Couteau, and C.~Soci,
  ``Gaas/algaas nanowire photodetector,'' \emph{Nano letters}, vol.~14, no.~5,
  pp. 2688--2693, 2014.

\bibitem{wang2013high}
H.~Wang, ``High gain single gaas nanowire photodetector,'' \emph{Applied
  Physics Letters}, vol. 103, no.~9, p. 093101, 2013.

\bibitem{garnett2011nanowire}
E.~C. Garnett, M.~L. Brongersma, Y.~Cui, and M.~D. McGehee, ``Nanowire solar
  cells,'' \emph{Annual Review of Materials Research}, vol.~41, pp. 269--295,
  2011.

\bibitem{kayes2008radial}
B.~M. Kayes, M.~Filler, M.~Henry, J.~Maiolo~Iii, M.~Kelzenberg, M.~Putnam,
  J.~Spurgeon, K.~Plass, A.~Scherer, N.~Lewis \emph{et~al.}, ``Radial pn
  junction, wire array solar cells,'' in \emph{2008 33rd IEEE Photovoltaic
  Specialists Conference}.\hskip 1em plus 0.5em minus 0.4em\relax IEEE, 2008,
  pp. 1--5.

\bibitem{hochbaum2010semiconductor}
A.~I. Hochbaum and P.~Yang, ``Semiconductor nanowires for energy conversion,''
  \emph{Chemical reviews}, vol. 110, no.~1, pp. 527--546, 2010.

\bibitem{sandhu2014detailed}
S.~Sandhu, Z.~Yu, and S.~Fan, ``Detailed balance analysis and enhancement of
  open-circuit voltage in single-nanowire solar cells,'' \emph{Nano letters},
  vol.~14, no.~2, pp. 1011--1015, 2014.

\bibitem{sun2011compound}
K.~Sun, A.~Kargar, N.~Park, K.~N. Madsen, P.~W. Naughton, T.~Bright, Y.~Jing,
  and D.~Wang, ``Compound semiconductor nanowire solar cells,'' \emph{IEEE
  Journal of Selected Topics in Quantum Electronics}, vol.~17, no.~4, pp.
  1033--1049, 2011.

\bibitem{lapierre2013iii}
R.~LaPierre, A.~Chia, S.~Gibson, C.~Haapamaki, J.~Boulanger, R.~Yee,
  P.~Kuyanov, J.~Zhang, N.~Tajik, N.~Jewell \emph{et~al.}, ``Iii--v nanowire
  photovoltaics: review of design for high efficiency,'' \emph{physica status
  solidi (RRL)--Rapid Research Letters}, vol.~7, no.~10, pp. 815--830, 2013.

\bibitem{kupec2009dispersion}
J.~Kupec and B.~Witzigmann, ``Dispersion, wave propagation and efficiency
  analysis of nanowire solar cells,'' \emph{Optics express}, vol.~17, no.~12,
  pp. 10\,399--10\,410, 2009.

\bibitem{kupec2010light}
J.~Kupec, R.~L. Stoop, and B.~Witzigmann, ``Light absorption and emission in
  nanowire array solar cells,'' \emph{Optics express}, vol.~18, no.~26, pp.
  27\,589--27\,605, 2010.

\bibitem{yoshimura2013indium}
M.~Yoshimura, E.~Nakai, K.~Tomioka, and T.~Fukui, ``Indium phosphide
  core--shell nanowire array solar cells with lattice-mismatched window
  layer,'' \emph{Applied Physics Express}, vol.~6, no.~5, p. 052301, 2013.

\bibitem{wen2011theoretical}
L.~Wen, Z.~Zhao, X.~Li, Y.~Shen, H.~Guo, and Y.~Wang, ``Theoretical analysis
  and modeling of light trapping in high efficicency gaas nanowire array solar
  cells,'' \emph{Applied Physics Letters}, vol.~99, no.~14, p. 143116, 2011.

\bibitem{li2015influence}
Z.~Li, Y.~C. Wenas, L.~Fu, S.~Mokkapati, H.~H. Tan, and C.~Jagadish,
  ``Influence of electrical design on core--shell gaas nanowire array solar
  cells,'' \emph{IEEE Journal of Photovoltaics}, vol.~5, no.~3, pp. 854--864,
  2015.

\bibitem{wu2018optimization}
Y.~Wu, X.~Yan, W.~Wei, J.~Zhang, X.~Zhang, and X.~Ren, ``Optimization of gaas
  nanowire pin junction array solar cells by using algaas/gaas
  heterojunctions,'' \emph{Nanoscale research letters}, vol.~13, no.~1, pp.
  1--7, 2018.

\bibitem{huang2012broadband}
N.~Huang, C.~Lin, and M.~L. Povinelli, ``Broadband absorption of semiconductor
  nanowire arrays for photovoltaic applications,'' \emph{Journal of Optics},
  vol.~14, no.~2, p. 024004, 2012.

\bibitem{joyce2013electronic}
H.~J. Joyce, C.~J. Docherty, Q.~Gao, H.~H. Tan, C.~Jagadish, J.~Lloyd-Hughes,
  L.~M. Herz, and M.~B. Johnston, ``Electronic properties of gaas, inas and inp
  nanowires studied by terahertz spectroscopy,'' \emph{Nanotechnology},
  vol.~24, no.~21, p. 214006, 2013.

\bibitem{bai2014one}
F.~Bai, M.~Li, R.~Huang, Y.~Li, M.~Trevor, and K.~P. Musselman, ``A one-step
  template-free approach to achieve tapered silicon nanowire arrays with
  controllable filling ratios for solar cell applications,'' \emph{RSC
  Advances}, vol.~4, no.~4, pp. 1794--1798, 2014.

\bibitem{zhang2018photovoltaic}
J.~Zhang, L.~Ai, X.~Yan, Y.~Wu, W.~Wei, M.~Zhang, and X.~Zhang, ``Photovoltaic
  performance of pin junction nanocone array solar cells with enhanced
  effective optical absorption,'' \emph{Nanoscale research letters}, vol.~13,
  no.~1, pp. 1--9, 2018.

\bibitem{anttu2010coupling}
N.~Anttu and H.~Xu, ``Coupling of light into nanowire arrays and subsequent
  absorption,'' \emph{Journal of nanoscience and nanotechnology}, vol.~10,
  no.~11, pp. 7183--7187, 2010.

\bibitem{zhan2014enhanced}
Y.~Zhan, X.~Li, S.~Wu, K.~Li, Z.~Yang, and A.~Shang, ``Enhanced photoabsorption
  in front-tapered single-nanowire solar cells,'' \emph{Optics letters},
  vol.~39, no.~19, pp. 5756--5759, 2014.

\bibitem{soci2008systematic}
C.~Soci, X.-Y. Bao, D.~P. Aplin, and D.~Wang, ``A systematic study on the
  growth of gaas nanowires by metal- organic chemical vapor deposition,''
  \emph{Nano letters}, vol.~8, no.~12, pp. 4275--4282, 2008.

\bibitem{chuang2011GaAs}
L.~C. Chuang, M.~Moewe, K.~W. Ng, T.-T.~D. Tran, S.~Crankshaw, R.~Chen, W.~S.
  Ko, and C.~Chang-Hasnain, ``Gaas nanoneedles grown on sapphire,''
  \emph{Applied Physics Letters}, vol.~98, no.~12, p. 123101, 2011.

\bibitem{lapierre2011numerical}
R.~LaPierre, ``Numerical model of current-voltage characteristics and
  efficiency of gaas nanowire solar cells,'' \emph{Journal of Applied Physics},
  vol. 109, no.~3, p. 034311, 2011.

\bibitem{jiang2012long}
N.~Jiang, P.~Parkinson, Q.~Gao, S.~Breuer, H.~Tan, J.~Wong-Leung, and
  C.~Jagadish, ``Long minority carrier lifetime in au-catalyzed gaas/alxga1-
  xas core-shell nanowires,'' \emph{Applied Physics Letters}, vol. 101, no.~2,
  p. 023111, 2012.

\bibitem{aaberg2015GaAs}
I.~{\AA}berg, G.~Vescovi, D.~Asoli, U.~Naseem, J.~P. Gilboy, C.~Sundvall,
  A.~Dahlgren, K.~E. Svensson, N.~Anttu, M.~T. Bj{\"o}rk \emph{et~al.}, ``A
  gaas nanowire array solar cell with 15.3\% efficiency at 1 sun,'' \emph{IEEE
  Journal of Photovoltaics}, vol.~6, no.~1, pp. 185--190, 2015.

\bibitem{demichel2010impact}
O.~Demichel, M.~Heiss, J.~Bleuse, H.~Mariette, and A.~Fontcuberta~i Morral,
  ``Impact of surfaces on the optical properties of gaas nanowires,''
  \emph{Applied Physics Letters}, vol.~97, no.~20, p. 201907, 2010.

\bibitem{chang2012electrical}
C.-C. Chang, C.-Y. Chi, M.~Yao, N.~Huang, C.-C. Chen, J.~Theiss, A.~W.
  Bushmaker, S.~LaLumondiere, T.-W. Yeh, M.~L. Povinelli \emph{et~al.},
  ``Electrical and optical characterization of surface passivation in gaas
  nanowires,'' \emph{Nano letters}, vol.~12, no.~9, pp. 4484--4489, 2012.

\bibitem{mariani2013GaAs}
G.~Mariani, A.~C. Scofield, C.-H. Hung, and D.~L. Huffaker, ``Gaas
  nanopillar-array solar cells employing in situ surface passivation,''
  \emph{Nature communications}, vol.~4, no.~1, pp. 1--8, 2013.

\bibitem{yao2014GaAs}
M.~Yao, N.~Huang, S.~Cong, C.-Y. Chi, M.~A. Seyedi, Y.-T. Lin, Y.~Cao, M.~L.
  Povinelli, P.~D. Dapkus, and C.~Zhou, ``Gaas nanowire array solar cells with
  axial p--i--n junctions,'' \emph{Nano letters}, vol.~14, no.~6, pp.
  3293--3303, 2014.

\bibitem{shur1990physics}
M.~Shur, ``Physics of semiconductor devices, prentice hall,'' \emph{Inc.,
  Englewood Cliffs, New Jersey}, p. 680, 1990.

\bibitem{charge}
``Lumerical charge solver,''
  \url{https://support.lumerical.com/hc/en-us/articles/360034917693-CHARGE-solver-introduction},
  April 2021.

\bibitem{fdtd}
``Lumerical fdtd solver,''
  \url{https://support.lumerical.com/hc/en-us/articles/360034914633-Finite-Difference-Time-Domain-FDTD-solver-introduction},
  April 2021.

\bibitem{molenkamp1988very}
L.~Molenkamp and H.~Van’t~Blik, ``Very low interface recombination velocity
  in (al, ga) as heterostructures grown by organometallic vapor-phase
  epitaxy,'' \emph{Journal of applied physics}, vol.~64, no.~8, pp. 4253--4256,
  1988.

\bibitem{wu2017efficient}
D.~Wu, X.~Tang, K.~Wang, Z.~He, and X.~Li, ``An efficient and effective design
  of inp nanowires for maximal solar energy harvesting,'' \emph{Nanoscale
  research letters}, vol.~12, no.~1, pp. 1--10, 2017.

\bibitem{fossum1982physical}
J.~Fossum and D.~Lee, ``A physical model for the dependence of carrier lifetime
  on doping density in nondegenerate silicon,'' \emph{Solid-State Electronics},
  vol.~25, no.~8, pp. 741--747, 1982.

\bibitem{fossum1983carrier}
J.~Fossum, R.~Mertens, D.~Lee, and J.~Nijs, ``Carrier recombination and
  lifetime in highly doped silicon,'' \emph{Solid-State Electronics}, vol.~26,
  no.~6, pp. 569--576, 1983.

\bibitem{tajik2011sulfur}
N.~Tajik, Z.~Peng, P.~Kuyanov, and R.~LaPierre, ``Sulfur passivation and
  contact methods for gaas nanowire solar cells,'' \emph{Nanotechnology},
  vol.~22, no.~22, p. 225402, 2011.

\bibitem{jung2010strong}
J.-Y. Jung, Z.~Guo, S.-W. Jee, H.-D. Um, K.-T. Park, and J.-H. Lee, ``A strong
  antireflective solar cell prepared by tapering silicon nanowires,''
  \emph{Optics Express}, vol.~18, no. 103, pp. A286--A292, 2010.

\bibitem{chattopadhyay2010anti}
S.~Chattopadhyay, Y.-F. Huang, Y.-J. Jen, A.~Ganguly, K.~Chen, and L.~Chen,
  ``Anti-reflecting and photonic nanostructures,'' \emph{Materials Science and
  Engineering: R: Reports}, vol.~69, no. 1-3, pp. 1--35, 2010.

\end{thebibliography}
\bibliographystyle{IEEEtran}

\end{document}